# Stability of an Exciton bound to an Ionized Donor in Quantum Dots

by

S. Baskoutas[1*)], W. Schommers[2)], A. F. Terzis[3)], V. Kapaklis[4)], M. Rieth[5)], C. Politis[4,6)]

[1)]Materials Science Department, University of Patras, 26500 Patras, Greece.
[2)] Forschungszentrum Karlsruhe, Hauptabteilung Informations-und Kommunikationstechnik, 76021 Karlsruhe, Germany
[3)] Physics Department, University of Patras, 26500 Patras, Greece.
[4)]Engineering Science Department, University of Patras, 26500 Patras, Greece.
[5)] Forschungszentrum Karlsruhe, Institut für Materialforschung I, POBox 3640, 76021 Karlsruhe, Germany
[6)] Forschungszentrum Karlsruhe, Institut für Nanotechnologie, Forschungszentrum Karlsruhe, 76021 Karlsruhe, Germany

## Abstract
Total energy, binding energy, recombination rate (of the electron hole pair) for an exciton (X) bound in a parabolic two – dimensional quantum dot by a donor impurity located on the z – axis at a distance $d$ from the dot plane, are calculated by using the Hartree formalism with a recently developed numerical method (PMM) for the solution of the Schrödinger equation. As our analysis indicates there is a critical dot radius $R^c$ such that for $R < R^c$ the complex is unstable and with an increase of the impurity distance this critical radius increases. Furthermore, there is a critical value of the mass ratio $\sigma = m_e^* / m_h^*$ such that for $\sigma < \sigma^c$ the complex is stable. The appearance of this stability condition depends both on the impurity distance and the dot radius, in a way that with an increase of the impurity distance we have an increase in the maximum dot radius where this stability condition appears. For dot radii greater than this maximum dot radius (for fixed impurity distance) the complex is always stable.

---

*Corresponding author: S. Baskoutas
e-mail: bask@des.upatras.gr , Tel.: +30610996370, Fax: +30610996370

## 1. Introduction

The three-particle complex (D$^+$, X) consisting of an electron and a hole bound to an ionized donor is the simplest possible bound – exciton complex. Since Lampert [1] first suggested the possibility of the binding of an exciton (X) to neutral (D$^0$) and ionized (D$^+$) shallow impurities in semiconductors, a great deal of effort has been devoted to the study of these compounds both in bulk (3D) [2] and confined compound semiconductors [3].

In bulk semiconductors, it has been proven [4] that the (D$^+$, X) complex is only stable when the electron to hole effective mass ratio $\sigma = m_e^* / m_h^*$ is less than the critical value $\sigma_c^{3D} = 0.426$. However, more favorable conditions for the detection of the (D$^+$, X) complex are expected [5,6] for lower dimension systems.

In the present paper we concentrate our study on the stability conditions of the barrier (D$^+$, X) complex in a quantum dot (QD) with parabolic confinement, using the Hartree formalism with a recently developed numerical method for the solution of the Schrödinger equation, which is called Potential Morphing Method PMM [7,8]. We calculate the total energy of the complex, the ground state energy of the neutral donor, the binding energy of the complex and the recombination rate (wavefunction overlapping) of the electron – hole pair for various dot lengths, various rates $\sigma$ and different impurity positions along the z–axis. In our analysis we assume also that the considered quantum dot system is a two – dimensional system, where its height $\Delta z \ll R$ ($R$ is the radius of the dot) [9]. Therefore, in Sec. 2 we present the calculation method and we obtain the total energy of the complex, the binding energy as well as the wavefunction overlapping (recombination rate). In Sec. 3 we present numerical results and we discuss them. Finally Sec. 4 is devoted to concluding remarks.

## 2. Theory

### a. Binding Energy

Within the effective – mass approximation, the Hamiltonian of the (D$^+$,X) complex in a QD, consisting of an electron and hole bound to ionized donor impurity (D$^+$) located at a fixed distance $d$ along the z–axis, subjected to a parabolic confinement potential can be written as [10]

$$H = \frac{p_e^2}{2m_e^*} + \frac{1}{2}m_e^*\omega_0^2 r_e^2 + \frac{p_h^2}{2m_h^*} + \frac{1}{2}m_h^*\omega_0^2 r_h^2 + V_c(r_e) + V_c(r_h) + V_c(r_e, r_h) \quad (1)$$

with

$$V_c(r_e) = -\frac{e^2}{\varepsilon\sqrt{r_e^2 + d^2}} \qquad (2)$$

$$V_c(r_h) = \frac{e^2}{\varepsilon\sqrt{r_h^2 + d^2}} \qquad (3)$$

and [11]

$$V_c(r_e, r_h) = -e^2 \big/ \varepsilon \big(|r_e - r_h|^2 + \Delta z^2\big)^{1/2} + e^2 \big/ \varepsilon \big(|r_e - r_h|^2 + (\Delta z + \delta)^2\big)^{1/2} \quad (4)$$

where $\varepsilon$ is the static dielectric constant and $\omega_0$ is the strength of the confinement. Furthermore, the first term in eq. (4) corresponds to the electron – hole interaction with a cut - off at the finite extent of the wavefunction $\Delta z$ along the z–axis (which denotes a very strong confinement along the z–axis) and the second term corresponds to the electron – induced hole charge interaction with a cut - off at $\Delta z + \delta$, where $\delta$ is the width of an interface surrounding the dot and which tends to zero for infinite confinement [8,12].

The eigenfunctions of the Hamiltonian (1) are postulated in the Hartree form

$$\Psi(r_e, r_h) = \Phi_e(r_e)\Phi_h(r_h) \qquad (5)$$

In such an approximation the equation for the eigenstates of (1) is equivalent to the following system of equations

$$\left[\frac{p_i^2}{2m_i^*} + U_i(r_i)\right]\Phi_i(r_i) = E_i\Phi_i(r_i) \qquad (i = e, \ h) \qquad (6)$$

where the self – consistent effective field that acts on the electron, is given by the formula

$$U_e(r_e) = V_C(r_e) + \frac{1}{2}m\omega_0^2 r_e^2 - \frac{e^2}{\varepsilon}\int dr_h \frac{|\Phi_h(r_h)|^2}{\left(|r_e - r_h|^2 + \Delta z^2\right)^{1/2}} + \frac{e^2}{\varepsilon}\int dr_h \frac{|\Phi_h(r_h)|^2}{\left(|r_e - r_h|^2 + (\Delta z + \delta)^2\right)^{1/2}}$$

$$(7)$$

and the self - consistent effective field that acts on the hole $U_h(r_h)$ is obtained by changing the indices in the above formula: e into h and vice versa.

In order to find eigenvalues and eigenfunctions of eq. (6) with the PMM [7,8] we need a reference system with well known eigenfunctions and eigenvalues, as for example the simple harmonic oscillator in 2 – dimensions ($U_{HO}(r_i)$). The essential point now is that the transition from the known system to the unknown system ($U_i(r_i)$) can be performed by means of the time – dependent Schrödinger equation

$$i\hbar\,\partial\Phi_i(r_i,t)/\partial t = \left[\frac{p_i^2}{2m_i^*} + U_t(r_i)\right]\Phi_i(r_i,t) \tag{8}$$

with

$$U_t(r_i) = \sigma(t)\,U_i(r_i) + [1-\sigma(t)]\,U_{HO}(r_i) \tag{9}$$

where $\sigma(t)$ has the property: $\sigma(t)=0$, for $t \le t_a$ and $\sigma(t)=1$ for $t \ge t_b$. For $t_a \le t \le t_b$ the function $\sigma(t)$ may have any shape but should increase monotonically. A simple choice which we have used in our calculations in the present paper is $\sigma(t)=a(t-t_a)$ with $a=1/(t_b-t_a)$.

For $t \ge t_b$ the $\Phi_i(r_i)$ of eq. (6) is given by

$$\Phi_i(r_i) = \Phi_i(r_i,t_b) \tag{10}$$

and the energy $E_i$ (eq. 6) is obtained, using eq. (10) as follows

$$E_i = \int_{-\infty}^{\infty}\Phi_i^*(r_i)\left[\frac{p_i^2}{2m_i^*} + U_i(r_i)\right]\Phi_i(r_i)dr_i \tag{11}$$

Then the total energy of the (D$^+$,X) complex after the convergence in the Hartree scheme has the form

$$E(D^+,X) = E_e + E_h \qquad (12)$$

and the binding energy is [10]

$$E_b = E(D^0) + E_h - E(D^+,X) \qquad (13)$$

where $E(D^0)$ is the ground state energy of the neutral donor and $E_h$ is the lowest level of a hole in the QD without the Coulomb potential. According to the corresponding stability conditions [10,13], the (D$^+$,X) complex is stable when $E_b > 0$.

b. *Recombination rate*

The wavefunction overlapping is defined as follows [14]

$$f_{eh} = \left| \int \Phi_e^*(r_e) \Phi_h(r_h) dr_e dr_h \right|^2 \qquad (14)$$

where the above wavefunctions of the electron and hole are obtained within the PMM procedure (eq. (10)), after the convergence in the Hartree scheme.

Assuming that the decay time is inversely proportional to the recombination rate [14], we can study the decay behavior of the e – h pair in the presence of the impurity charge.

## 3. Results

Solving the system of equations (6) self – consistently in the Hartree scheme we take as initial wavefunctions (eq. (5)) these of the usual 2D – harmonic oscillator [15]. Every time we need a solution of (6), we use the PMM scheme [7,8] with reference system that of the 2D – harmonic oscillator, and time interval $\Delta t = 10^{-4}$ (in units $m_i^* R^2 / \hbar$ where $R$ is in m) and dx = dy = 0.025 (in units of $R$ (nm)). In every iteration step of the Hartree scheme we obtain with PMM both the energies and the corresponding wavefunctions which are normalized. The convergence in the self – consistent Hartree procedure is obtained in three or five rounds.

We consider the case of GaAs semiconductor and we choose as material parameters $\varepsilon$ = 12.4, $m_e^*$ =0.067 $m_e$ , $\sigma$ = 0.707 [10] and $\Delta z$ = 0.2$R$ [8,9]) and $\delta$ = $10^{-5}R$ [8][*]. The fixed $\delta$ value is specified as the distance below which the mean value of the Coulomb potential $\langle V_c(r_e, r_h) \rangle$ does not change. Obviously, for very low $\delta$ values the mean value of the potential gives unphysical results due to computational round off errors. Furthermore, for given value of $\Delta z$ ($\Delta z$ <<$R$) corresponds a value of $\delta$ below which the mean value of the Coulomb potential $\langle V_c(r_e, r_h) \rangle$ is the same as above. Working in dimensionless units we define the radius of the QD as [8] $R = (\hbar/m\omega)^{1/2}$ , and we calculate $E(D^+, X)$ as a function of the QD radius for various positions of the impurity along the z – axis. As is clearly depicted in Fig. 1, $E(D^+, X)$ decreases with the increase of $R$, but takes larger values as the impurity moves away from the QD plane. This result, which is in agreement with the corresponding result of ref. [14], essentially denotes that the smaller the radius of the dot is and/or the larger the impurity distance from the dot plane is, the higher the difficulty for the exciton to bound to the charged impurity. Furthermore in Fig. 2 we depict the behavior of the neutral donor ground state energy $E(D^0)$, as a function of the QD radius. As is clearly seen $E(D^0)$ increases as the dot radius decreases (confinement effect) [16]. It is also obvious that as the impurity moves away from the QD center $E(D^0)$ takes larger values. Furthermore in Fig. 3 we present the behavior of the binding energy $E_b$ as a function of the QD radius. As is clearly seen there is a critical value $R^c$, such that for $R < R^c$ the (D$^+$, X) complex becomes unstable [10]. This critical value depends strongly on the position of the impurity and as we can see from Fig. 3, $R^c$ increases when the impurity distance from the QD plane increases.

---

[*]The same value of $\delta$ has been obtained in a similar two dimensional quantum dot system [8] and due to the large confinement along the z – axis, we expect that it tends to zero [12].

Now in order to see the decay behavior of the e – h pair in the presence of the impurity, we have plotted in Fig. 4 the wavefunction overlapping $f_{eh}$ as a function of the QD radius for various impurity positions along the z – axis. Taking into account that the decay time is inversely proportional to the wavefunction overlapping, we can see that with the increase of the QD radius the wavefunction overlapping decreases – the decay time increases e.g. the electron – hole pair is mole stable. Furthermore, $f_{eh}$ takes larger values as the distance of the impurity from the QD plane increases. This result which is in agreement with the corresponding result of ref. [14], is in agreement also with the results depicted in Figs. 1 and 3, taking into account that larger $E\left(D^+, X\right)$ means larger difficulty for the exciton to bound to the charged impurity and smaller $E_b$ means that the hole is more free to move away from the complex and thus the system is more unstable and vice versa.

In Fig. 5 we have plotted the behavior of the binding energy as a function of the mass ratio $\sigma$ for $R$=7 nm and the result we have obtained points out that there is a critical value $\sigma^c$ ($0.426 = \sigma_{3D}{}^c < \sigma^c < \sigma_{2D}{}^c = 0.88$ [13]) such that for $\sigma < \sigma^c$ the system is stable. The appearance of this behavior depends from the impurity distance as well as from the dot radius. As is seen with the increase of the impurity distance $\sigma^c$ decreases and also increases the maximum dot radius in which this behavior appears. For dot radii greater than this maximum radius (for fixed impurity distance) the complex is stable (Fig. 6). Finally, in Fig. 7 we depict the behavior of the wavefunction overlapping as function of the mass ratio $\sigma$ for $R$=7 nm and we see that as $\sigma$ increases $f_{eh}$ increases and furthermore as the distance of the impurity increases, $f_{eh}$ also increases – which is in agreement with Fig. 5.

## 4. Conclusions

In the present work we have studied the stability of $(D^+, X)$ complex in a two – dimensional GaAs QD with a height $\Delta z << R$. Using the Hartree formalism with PMM for the solution of the Schrödinger equation, we have obtained the total energy of the complex $E\left(D^+, X\right)$, the binding energy $E_b$ and the recombination rate $f_{eh}$ of the electron hole pair. As our results indicate there is a critical radius $R^c$, such that for $R < R^c$ the $(D^+, X)$ complex becomes unstable. This critical radius depends on the impurity distance from the dot plane in a way that with an increase of the impurity distance the critical radius increases. This behavior as regards the stability is in agreement with the behavior of the recombination rate of the electron hole pair, which decreases with the increase of $R$ and also takes larger values as the impurity distance from the dot plane increases. Finally, for a fixed dot radius ($R$=7nm) we have obtained a critical value of the mass ratio $\sigma^c$, such that for $\sigma < \sigma^c$ the system is stable. As is seen with the increase of the impurity distance $\sigma^c$ decreases and certainly increases the maximum dot radius in which this stability condition appears. For dot radii greater than this maximum radius (for fixed impurity distance) the complex is always stable.

**Acknowledgements**

This work has been supported by the cooperation between the Forschungszentrum Karlsruhe, Germany and the University of Patras, Greece and also by the International Office of the Bundesministerium für Bildung und Forschung, Germany. The authors would like to thank Mr. Frank Schmitz, from Hauptabteilung Informations-und Kommunikationstechnik (HIK), Forschungszentrum Karlsruhe, Germany for his valuable help with the VPP Supercomputer. The authors S. B., V. K., and C. P., would like to thank also the Research Committee of University of Patras, Greece, for financial support under the project Karatheodoris: 'Synthesis, Characterization and Properties of Nanostructured Semiconductors'.

**Figure Captions**

**Fig. 1:** The energy $E(D^+, X)$ of the exciton complex as a function of the QD radius $R$, for various impurity distances ($d$) from the QD plane and for $\sigma = 0.707$.

**Fig. 2:** The neutral donor ground state energy $E(D^0)$ as a function of the QD radius $R$, for various impurity distances ($d$) from the QD plane and for $m_e^* = 0.067\, m_e$.

**Fig. 3:** The binding energy $E_b$ of the exciton complex as a function of the QD radius $R$, for various impurity distances ($d$) from the QD plane and for $\sigma = 0.707$. It is clearly seen the critical dot radius $R^c$ such that for $R < R^c$ the complex is unstable. Furthermore it is obvious that this critical dot radius increases as the impurity distance from the dot plane increases.

**Fig. 4:** The wavefunction overlapping $f_{\text{eh}}$ as a function of the QD radius $R$ for various impurity positions ($d$) along the z–axis and for $\sigma = 0.707$.

**Fig. 5:** The binding energy of the complex as a function of the mass ratio $\sigma = m_e^*/m_h^*$ for various impurity positions ($d$) along the z–axis and for $R=7$ nm. It is clearly seen the critical value $\sigma^c$ of the mass ratio for each impurity distance such that for $\sigma < \sigma^c$ the complex is stable. It is also obvious that this critical value $\sigma^c$ decreases as the impurity distance from the dot plane increases.

**Fig. 6:** The binding energy of the complex as a function of the mass ratio $\sigma = m_e^*/m_h^*$ for impurity distance $d=0.2$nm and $R=7$nm, $R=8$nm and $R=20$nm.

**Fig. 7:** The wavefunction overlapping $f_{\text{eh}}$ as a function of the mass ratio $\sigma$ for various impurity positions ($d$) along the z–axis and for $R=7$nm.

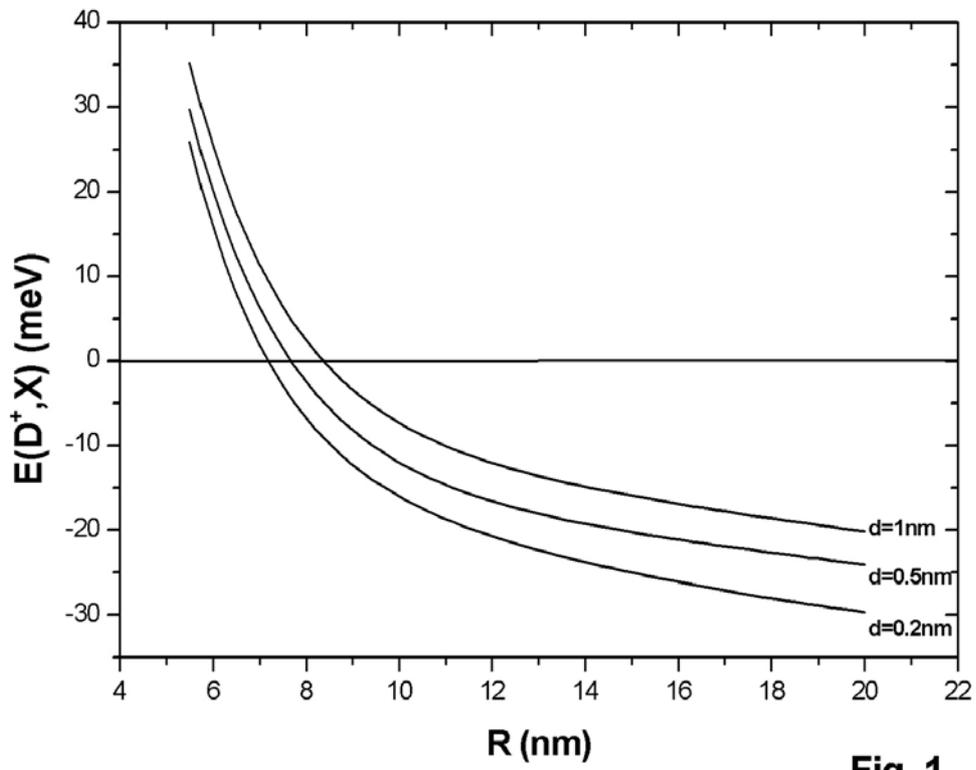

**Fig. 1**

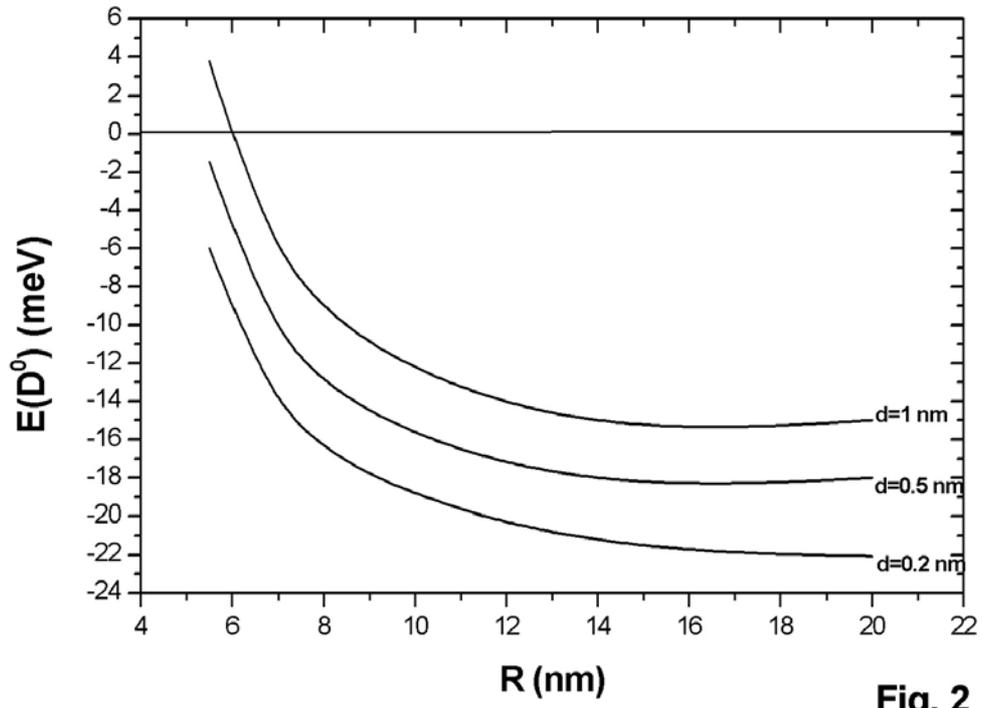

**Fig. 2**

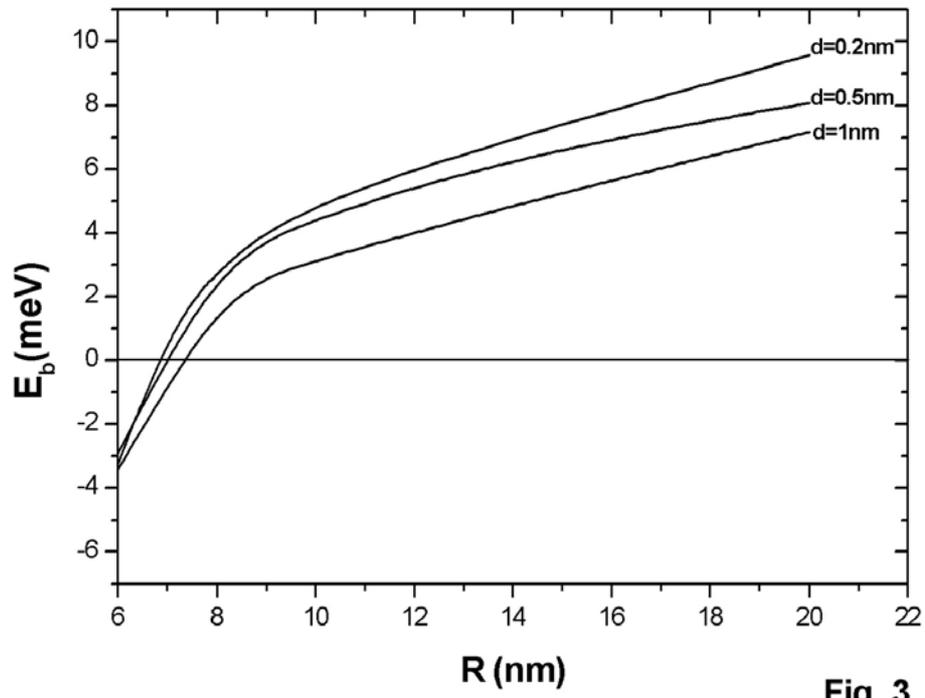

**Fig. 3**

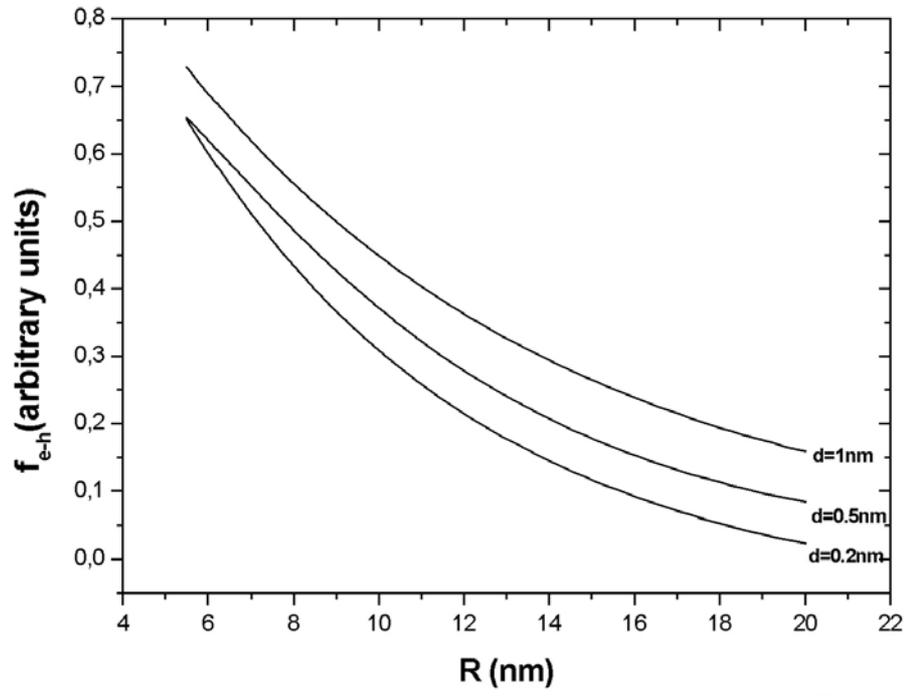

**Fig. 4**

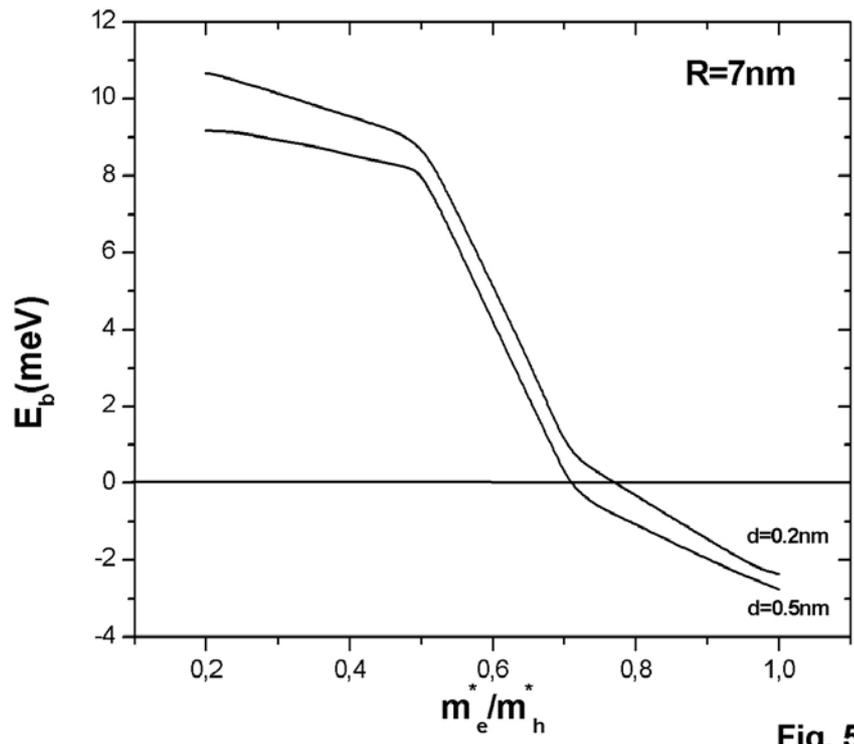

**Fig. 5**

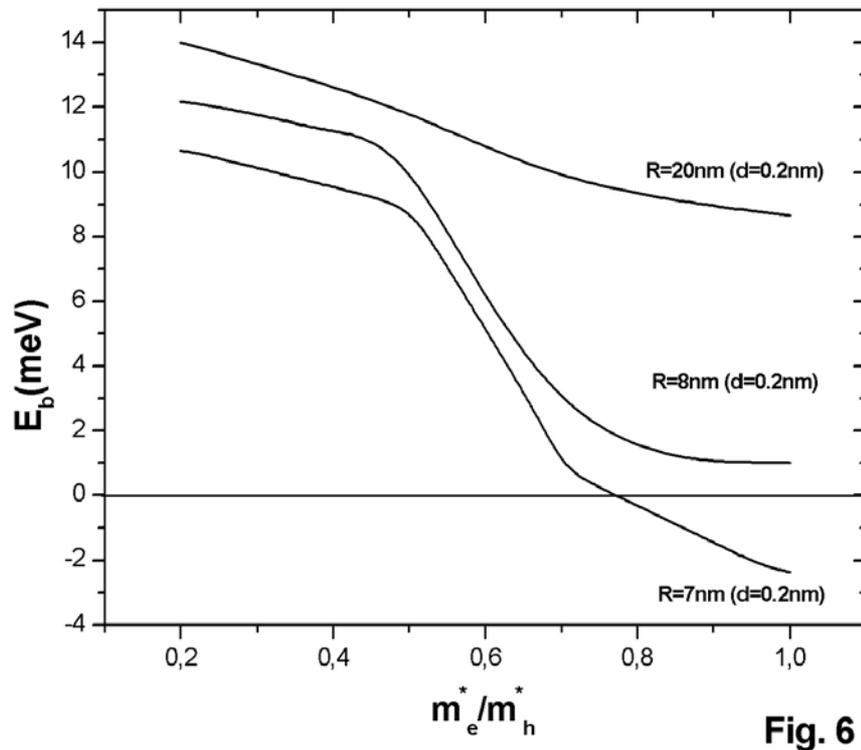

**Fig. 6**

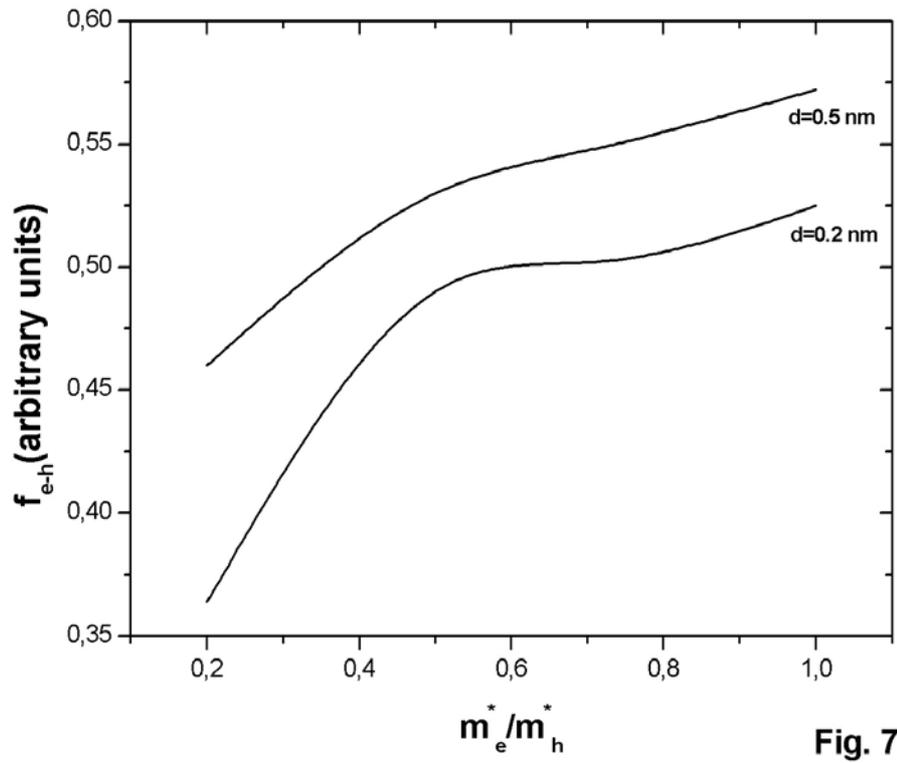

**Fig. 7**